\documentclass{emulateapj}
\shorttitle{Distant Giant Planet?}
\shortauthors{Lawler et al.}

\begin{document}

\title{Observational Signatures of a Massive Distant Planet on the Scattering Disk}

\author{S.~M.~Lawler\altaffilmark{1}, C.~Shankman\altaffilmark{2}, N.~Kaib\altaffilmark{3}, M.~T.~Bannister\altaffilmark{2}, B.~Gladman\altaffilmark{4}, J.J.~Kavelaars\altaffilmark{1,2}}

\altaffiltext{1}{National Research Council of Canada, Astronomy \& Astrophysics  Program,  5071  West  Saanich  Rd,  Victoria,  V9E  2E7, Canada}
\altaffiltext{2}{Department of Physics and Astronomy, University of Victoria, PO Box 1700, STN CSC Victoria, BC V8W 2Y2, Canada}
\altaffiltext{3}{HL Dodge Department of Physics \& Astronomy, University
of Oklahoma, Norman, OK 73019, USA}
\altaffiltext{4}{Department of Physics and Astronomy, University of British Columbia, 6224 Agricultural Road, Vancouver, BC V6T 1Z1, Canada}

\begin{abstract}

The orbital element distribution of trans-Neptunian objects (TNOs) with large pericenters has been suggested to be influenced by the presence of an undetected, large planet at $>$200 AU from the Sun. 
To find additional observables caused by this scenario, we here present the first detailed emplacement simulation in the presence of a massive ninth planet on the distant Kuiper Belt.
We perform 4~Gyr N-body simulations with the currently known Solar System planetary architecture, plus a 10~$M_{\oplus}$ planet with similar orbital parameters to those suggested by \citet{TrujilloSheppard2014} or \citet{BatyginBrown2016}, and 10$^5$ test particles in an initial planetesimal disk.
We find that including a distant superearth-mass planet produces a substantially different orbital distribution for the scattering and detached TNOs, raising the pericenters and inclinations of moderate semimajor axis ($50<a<500$~AU) objects.
We test whether this signature is detectable via a simulator with the observational characteristics of four precisely characterized TNO surveys.
We find that the qualitatively very distinct Solar System models that include a ninth planet are essentially observationally indistinguishable from an outer Solar System produced solely by the four giant planets.
We also find that the mass of the Kuiper Belt's current scattering and detached populations is required to be 3--10 times larger in the presence of an additional planet.
We do not find any evidence for clustering of orbital angles in our simulated TNO population.
Wide-field, deep surveys targeting inclined high-pericenter objects will be required to distinguish between these different scenarios. 

\end{abstract}

\section{Introduction}

The trans-Neptunian region contains far more structure than the originally hypothesized flat vestigial disk \citep{Edgeworth1949,Kuiper1951}.  
While the bulk of the Kuiper Belt's mass is contained in the classical belt, 
which has trans-Neptunian objects (TNOs) on fairly circular, low inclination orbits, TNOs on higher eccentricity orbits are plentiful. Resonant TNOs are protected from close Neptune encounters and can attain high eccentricity, allowing them to be more easily detected when near perihelion. Scattering TNOs often approach the Sun even more closely, as by definition they are required to have scattering encounters with Neptune or another giant planet \citep{Gladmanetal2008} and thus have pericenters in the giant planet region.  Though scattering TNOs can have very large semimajor axis orbits ($a \gg 50$~AU) and only make up $\sim$2\% of the Kuiper Belt's total population \citep{Petitetal2011}, their very high eccentricities boost detection rates, allowing detailed study of the population's characteristics \citep{Shankmanetal2013, Adamsetal2014, Shankmanetal2016}.
Detached TNOs make up a larger fraction of the Kuiper Belt total population \citep[$\sim$29\% for $D\gtrsim100$~km;][]{Petitetal2011}, but never approach Neptune closely enough to have their orbits affected by scattering encounters, and so are much harder to detect, due to their high pericenter distances and large semi-major axes.  

The history of the understanding of the $a>50$~AU population is an important context that frames both our current conception of these distant TNOs and their implications for an additional planet in this region.
The $a\simeq40-50$~AU low-$e$ Kuiper Belt initially seemed promising as the long-sought source of the Jupiter-Family comets (JFCs). 
However, once the population was observationally constrained, the estimated escape rate from that region was too low to allow it to serve as a JFC source \citep{Duncanetal1995}.
It became clear that no near-circular belt in the trans-Neptunian region {\it could} feed in JFCs without creating a scattering structure of large-$a$ TNOs once strong encounters with Neptune begin \citep{DuncanLevison1997}, and 
the discovery of the first member of this population, 1996 TL$_{66}$, was nearly simultaneous with this theory \citep{Luuetal1997}.
Dynamical simulations \citep{DuncanLevison1997} showed the surprising possibility that a non-negligible fraction ($\sim$1\%) of the initial planetesimal disk could still be in the scattering structure today; it need not be in steady state with an eroding main Kuiper Belt.
In this picture, the present-day structure of this scattering population is a band of objects with perihelia dominantly in the range $q\simeq35-39$~AU \citep{DuncanLevison1997,Trujilloetal2000,Morbidellietal2004,LykawkaMukai2007}, steadily decreasing in number as a function of semi-major axis.
TNOs are displaced outwards almost solely by gravitational interactions with Neptune; TNOs with $q<35$~AU are rapidly depleted, and thus relatively rare, while TNOs with $q>38$~AU are extremely rare.
The recognition that TNOs with $q>38$~AU existed, and in what must be great numbers \citep{Gladmanetal2002} led to the realization that the perihelion distribution must be extended to larger values \citep{Morbidellietal2004}.
The current terminology in the literature is to use the term `detached' for TNOs whose orbits are not today evolving due to Neptune encounters \citep{Gladmanetal2008}, and scattering for those which are. 
Unfortunately this does not correspond to a simple perihelion cut, although $q=37$~AU is sometimes used \citep{LykawkaMukai2007}.

The existence of the detached population requires some other major process, either historical or ongoing, to produce TNOs on these orbits.
Sedna \citep{Brownetal2004} and recently-discovered 2012~VP$_{113}$ \citep{TrujilloSheppard2014} are currently the highest-pericenter examples of detached TNOs. 
Possible explanations for the production of detached orbits include close stellar flybys \citep[e.g.][]{KenyonBromley2004,BrasserSchwamb2015}, changes in galactic tides caused by different Solar position within the Galaxy \citep[e.g.][]{Kaibetal2011}, ``rogue planets'' which were ejected early in the Solar System's history \citep[e.g.][]{GladmanChan2006} and undiscovered, additional planets \citep[e.g.][]{Gladmanetal2002,Brownetal2004,SoaresGomes2013}.
\citet{LykawkaMukai2008} suggest the presence of a distant Earth-mass planet to explain some of the structure of the Kuiper Belt, but one of their key arguments requires that there be no objects in distant Neptune mean-motion resonances.  
Several distant resonances, including the 3:1, 4:1, and 5:1, have been shown by recent surveys to be heavily populated \citep{Gladmanetal2012,Alexandersenetal2014,Pikeetal2015}.  

Limits exist on the presence of distant Solar System planets: analysis of data from the Wide-Field Infrared Survey Explorer (WISE) has shown that Jupiter-mass planets can be ruled out within 26,000~AU of the Sun \citep{Luhman2014}, though a superearth would be too faint in infrared wavelengths to be yet observed in the distant outer Solar System \citep[5--20~$M_{\oplus}$;][]{Fortneyetal2016}. 

\citet{TrujilloSheppard2014} have proposed a superearth on a circular orbit at roughly 250~AU to explain the apparent clustering in argument of pericenter ($\omega$) of a half-dozen detached TNOs with large perihelia.
The Kozai-Lidov effect and the inclination instability proposed by \citet{MadiganMcCourt2016} both demonstrate mechanisms for the clustering of $\omega$, but TNOs affected by either mechanism will continue to precess, and so naturally the orbits will separate over time.
This idea is modified and expanded upon by \citet{BatyginBrown2016}, who find that an eccentric superearth is capable of maintaining clustering among high-pericenter TNOs.

Both \citet{TrujilloSheppard2014} and \citet{BatyginBrown2016} rely on data from the Minor Planet Center (MPC) database, the repository of the orbital parameters for all known TNOs\footnote{As of 16 May 2016, this database contains 1986 TNOs, Centaurs and scattering objects, $\sim 1300$ of which have orbits known from observation on multiple oppositions.}, but which contains no information about observational parameters of the surveys in which these objects were discovered. 
The MPC TNOs are from a multitude of different surveys, which largely have unreported pointings, limiting magnitudes, detection efficiencies, and tracking efficiency post-discovery; this masks the true number of TNOs in different dynamical classes \citep{Kavelaarsetal2008}.
Biased sampling is particularly prone to affecting the discovery and recovery of the observed high $a/e$ population. 
An effect such as the apparent clustering of pericenters could be produced or significantly modified in non-intuitive ways \citep[see][for a discussion]{SheppardTrujillo2016}.

A survey with fully recorded observational biases can be properly debiased, giving the true numbers of objects required to exist in the unseen population in order to match the number of detections \citep{Jonesetal2006}. 
We therefore select a subset of the published wide-field surveys, permitting highly precise tests of the effects of observational bias on the observability of the distant-TNO orbital distributions.
Our test suite is an ensemble of four well-characterized surveys (Section~\ref{sec:surveys}): the Canada-France Ecliptic Plane Survey \citep[CFEPS;][]{Petitetal2011}, the HiLat Survey \citep{Petitetal2016}, the survey of \citet{Alexandersenetal2014}, and the first two sky blocks from the Outer Solar System Origins Survey \citep[OSSOS;][]{Bannisteretal2016}. 

Our goal is to see what dynamical signatures the addition of a superearth-scale planet generates within the the scattering and detached populations, and test this prediction against published, well-characterized surveys.
In this paper, we measure the effect a distant superearth would have on the orbital distribution of the high-$q$ ($q>37$~AU), moderate-$a$ ($50<a<500$~AU) component of the trans-Neptunian populations, using a detailed dynamical simulation containing many thousands of test particles.
We consider this population because it orbits beyond the dynamical dominance of Neptune, will be gravitationally sculpted by any potential ninth planet, and has still has pericenters within the detectable range of existing surveys. 

We show that although the differences between the intrinsic distribution of high-pericenter TNOs in models with and without a ninth planet are substantial, after observational biases are applied, the differences are currently indistinguishable.  
The fact that almost all known scattering objects have $q$=35--38~AU has tended to be viewed as confirmation of the baseline scattering scenario; our results show that the detection biases in the scattering population are so strong that the $q>38$~AU population could be numerous, but so weakly detectable that they are not represented in the observed sample.
Also using the survey simulator, we compare the predicted number of objects in the distant Solar System, and find that having an additional planet requires 3-10 times as many objects in the moderate-$a$ population.

\section{Orbital Integrations}

In order to make a realistic model of the distant TNOs as influenced by a possible superearth, we begin with the framework for building a scattering TNO and Oort Cloud model used by \citet{Shankmanetal2016}, which is a modified version of the model from \citet{Kaibetal2011}.  
Our three dynamical simulations begin with a hundred thousand massless test particles distributed from 4-40 AU, along with the 4 giant planets on their present-day orbits. The ``control'' sample is identical to that used by \citet{Shankmanetal2016}, while the other two simulations have an additional superearth with similar parameters to what was suggested by \citet{BatyginBrown2016} (eccentric P9: $M=10~M_{\oplus}$, $a=500$~AU, $e=0.5$, $i=5^{\circ}$), and in the interest of completeness, what was suggested by \citet{TrujilloSheppard2014} (circular P9: $M=10~M_{\oplus}$, $a=250$~AU, $e=0.0$, $i=5^{\circ}$).
These test particles and planets are evolved forward in time for 4~Gyr, under the influence of stellar flybys and Galactic tides \citep[for details, see][]{Kaibetal2011}. 
In order to ensure that the scattering and detached populations are not contaminated by the initial 4-40~AU disk, any objects that have $q>34$~AU and $a<42$~AU at 3.5~Gyr into the simulation are removed \citep[this is the same procedure used in][]{Shankmanetal2016}, as we are not here interested in the classical belt region.

What makes this simulation much more powerful than previous analyses is the sheer number of particles.  As a result, this dynamical simulation was computationally expensive to run. Previous integrations of Kuiper belt and Oort cloud formation were able to be sped up through a combination of adaptive timestepping and the exclusion of planetary perturbations on very distant particles \citep[e.g.,][]{KaibQuinnBrasser2011}. However, the inclusion of a distant ninth planet prevents this shortcut. Consequently, our integrations consumed over 10$^5$ core-hours.  

\begin{figure}
\centering
\includegraphics[scale=0.4]{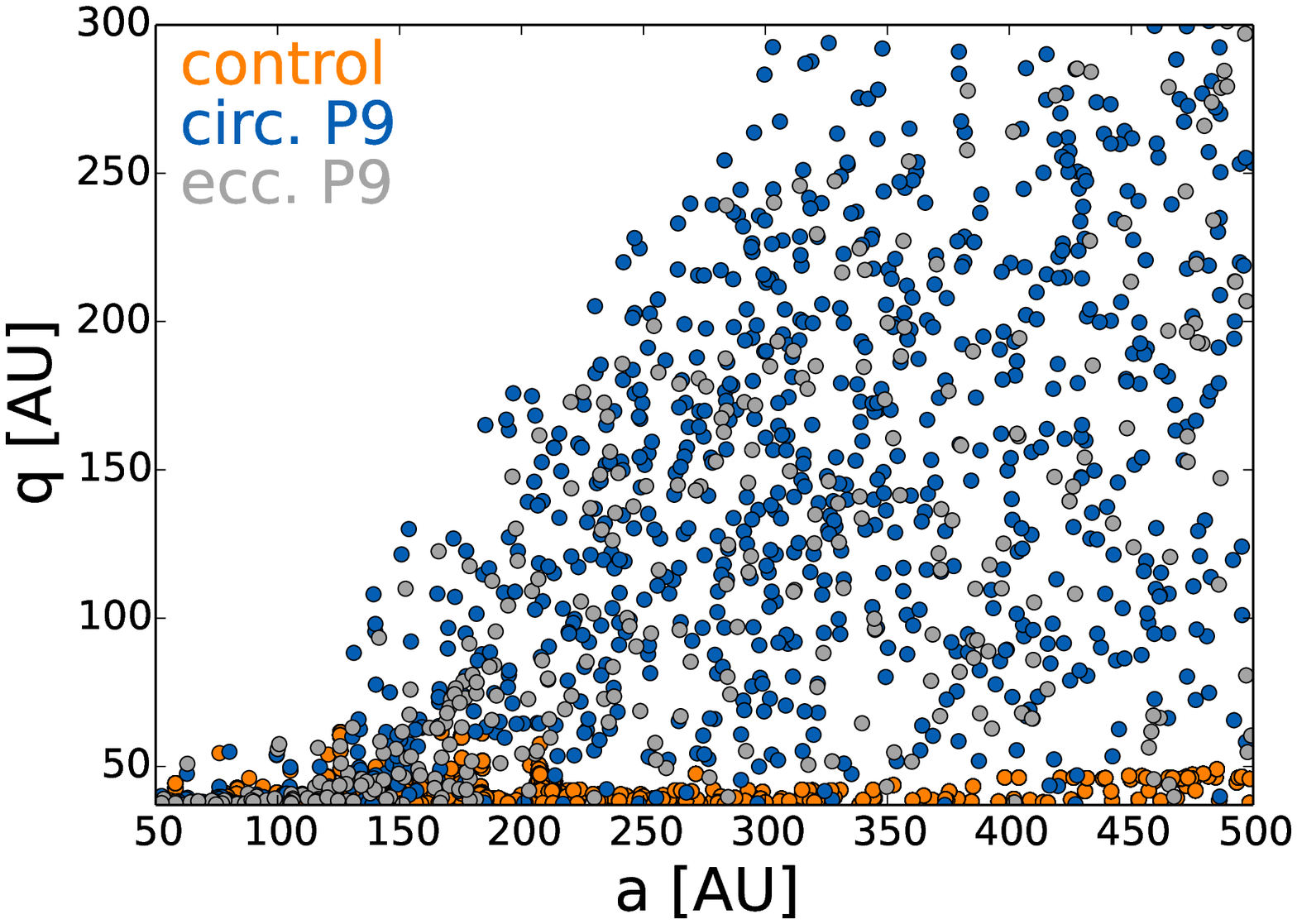}
\includegraphics[scale=0.4]{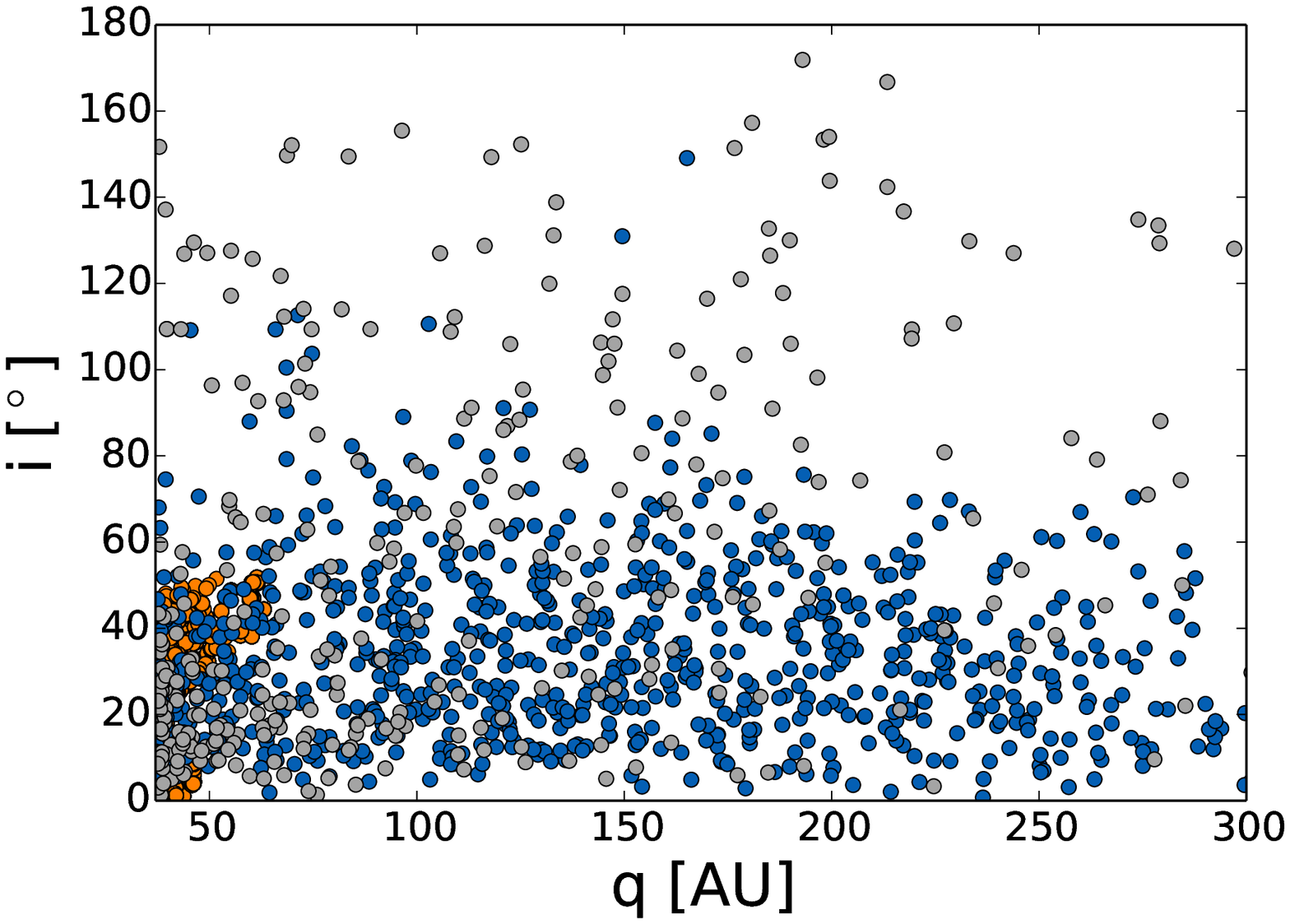}
\caption{
Orbital elements of all simulated TNOs with $q>37$~AU and $50<a<500$~AU from the control dynamical model (orange), the circular superarth dynamical model (blue), and the eccentric superearth dynamical model (gray).  
Left panel shows semi-major axis $a$ vs.\ pericenter distance $q$, right panel shows pericenter $q$ vs.\ inclination $i$.
The presence of a superearth on either a circular or eccentric orbit dramatically raises both the pericenter distribution and the inclination distribution of the distant TNOs.
}
\label{fig:aei}
\end{figure}

Figure~\ref{fig:aei} shows the orbital element distributions for the high-$q$ population in the control dynamical model \citep[the currently known Solar System;][]{Shankmanetal2016} and our nine-planet dynamical models after 4~Gyr of integration.  
The scattering TNO disk visible in the control dynamical model (orange) is the expected population of $q$=30-38~AU particles extending smoothly out to large $a$ (the classic scattering disk). 
At no semi-major axes (except a few rare resonant locations which can produce a few lower-$e$ particles via resonance sticking) do perihelia get raised into the detached region.
The introduction of a superearth results in frequent perihelion lifting for $a>150$~AU scattering objects, destroying the confinement and thus potentially offering a production method for the entire detached population all the way out to Sedna.

\begin{figure}
\centering
\includegraphics[scale=0.4]{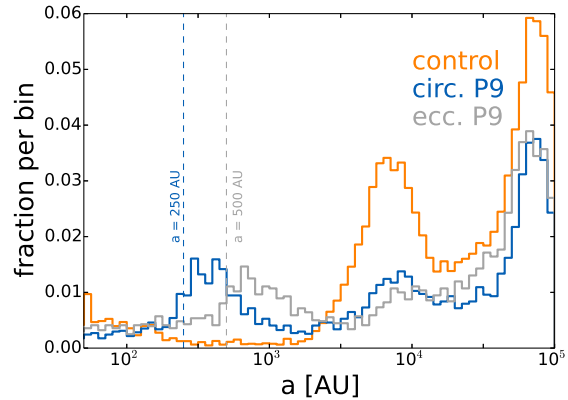}
\caption{
Semimajor axis distribution from 50 to 100,000~AU at the $\sim4.5$~Gyr state from three simulations: a control dynamical model (orange), the circular superearth dynamical model (blue), and the eccentric superearth dynamical model (gray).  
Histograms are normalized to the number of objects in each dynamical simulation.
The presence of a superearth on either a circular or eccentric orbit produces a concentration in objects with semimajor axes just outside that of the ninth planet. 
}
\label{fig:oort}
\end{figure}

Figure~\ref{fig:oort} shows the distribution of semimajor axes at the end of each of the three simulations.  
The control shows the number of moderate-$a$ objects per log($a$) bin steadily drops  with larger distances from Neptune; the scattering physics is poor at retaining
$a$=200-2000~AU objects over 4 Gyr \citep{Donesetal2004}.
Further out, there is a climb to a peak at the inner Oort Cloud, starting at roughly $a\sim2000$~AU as expected \citep{Gladmanetal2008}.  
Here the timescale of pericenter raising due to Galactic tides grows short enough that objects are efficiently decoupled from planetary scattering \citep{Duncanetal1987}.  
The most distant (and highest) peak is the main, or outer, Oort Cloud, where objects become isotropized by tidal torques.  
These two Oort Cloud peaks are also seen in the simulations with superearths.
Both P9 sims have an additional peak located just outside the semimajor axes of their respective ninth planets; in both simulations there is a $q\sim200$~AU peak, just inside $q$ of both superearths ($q=250$~AU). 

These simulated orbital distributions are very different from the control case with no additional planet; the next step is to determine whether or not these stark differences are observable with current surveys.

\section{Simulating Observations with Well-Calibrated Surveys} \label{sec:surveys}

We use the OSSOS survey simulator \citep{Bannisteretal2016, Shankmanetal2016}, which offers some improvements on the CFEPS survey simulator \citep{Jonesetal2006, Petitetal2011}.  
The survey simulator works by drawing objects from a dynamical model, applying survey biases for surveys where all the pointings, tracking efficiencies, and detection efficiencies are well-known, and determining whether or not a given simulated object could have been detected.  

When each object is drawn from our dynamical model, its major orbital elements ($a$, $e$, and $i$) are randomized within a small percentage of their model values, and its angular orbital elements ($\omega$, $\Omega$, and $\mathcal{M}$) are randomized.  
The object is also given an absolute $H_r$ magnitude using either the best-fit divot size distribution found by \citet{Shankmanetal2016} for the scattering population, or the knee size distribution preferred by \citet{Fraseretal2014},
but we find that this choice has no statistical effect on the analysis presented here.  
The object's simulated instantaneous on-sky position, 
distance, and resulting $r$-band magnitude determine whether or not this particular object would have been detected and tracked by any of the included surveys.
Simulated objects are drawn until the number of simulated detections specified by the user is met.

In this analysis, we use characterizations from four published surveys\footnote{Available for use as an ensemble at \url{http://dx.doi.org/10.5281/zenodo.31297}}.
A wide range of longitudes along the ecliptic are sampled by three surveys: CFEPS \citep{Petitetal2011}, \cite{Alexandersenetal2014}, and the OSSOS O and E blocks \citep{Bannisteretal2016}. 
High ecliptic latitudes are sampled by the HiLat survey \citep[]{Petitetal2016}.
We focus on the high-$q$, moderate-$a$ population ($q>37$~AU, $50$~AU~$<a<500$~AU). 
These are the objects most strongly perturbed by the distant superearth (Figure~\ref{fig:aei}). 
15 real TNOs have been detected in this $a$/$q$ cut in the above surveys, which allows estimation of absolute population numbers (Section~\ref{sec:pops}).

\subsection{Possible Superearth-Induced Structure in the Kuiper Belt Region Can't Yet be Observed} \label{sec:obs}

\begin{figure*}[h]
\centering
\includegraphics[scale=0.5]{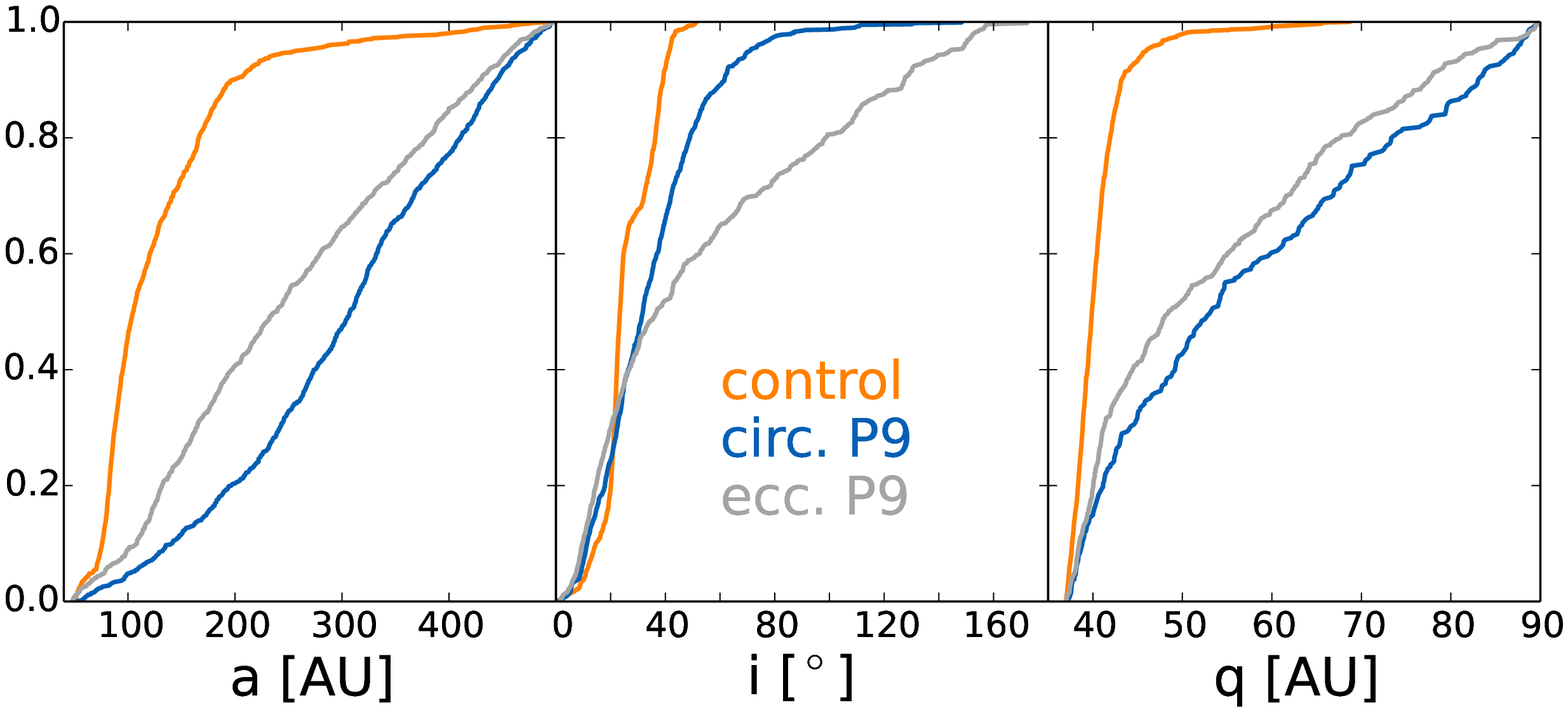}
\caption{
Cumulative histograms showing the intrinsic orbital distributions for moderate-$a$ orbits in each of the three dynamical models, to the same minimum and maximum values in each parameter: semi-major axis $a$, inclination $i$, and pericenter distance $q$.
The baseline Solar System dynamical model is shown in orange, the circular 9-planet dynamical model in blue and the eccentric 9-planet dynamical model in gray.  Only test particles with $q>37$~AU and $50<a<500$~AU are shown.
}
\label{fig:cumuhist_drawnonly}
\end{figure*}

Figure~\ref{fig:aei} highlights the differences in orbital elements of the high-$q$, moderate-$a$ population expected for no superearth, a circular superearth, and an eccentric superearth in the form of scatterplots.  
The very obvious difference is that the distant planet provides perturbations that raise inclinations and perihelia, potentially addressing two puzzles in Kuiper Belt science \citep{Gomes2003,Gladman2005}.
For comparative analysis, it is more straightforward to measure the differences between these distributions as cumulative distributions than scatterplots.
Figure~\ref{fig:cumuhist_drawnonly} shows a comparison between the three different dynamical models as cumulative distributions in three different orbital parameters: $a$, $i$, and $q$.  
Here each distribution has been cut at the same minimum and maximum values for each parameter. 

The strong differences between the three model distributions are immediately apparent in Figure~\ref{fig:cumuhist_drawnonly}.
Both superearth dynamical models result in more uniform distributions in $a$: $<10\%$ of the surviving high-$q$ population have orbits with $a<100$~AU, while in the control dynamical model about 50\% have orbits with $a<100$~AU.
The fraction of the intrinsic distributions with $a<100$~AU is a clear diagnostic of the presence of a distant superearth.
The control dynamical model has essentially no test particles with inclinations higher than $\sim$45$^{\circ}$, while the circular superearth dynamical model has $\sim$20\% of objects with $i>45^{\circ}$, and the eccentric superearth dynamical model has $\sim$40\% of objects with $i>45^{\circ}$ and $\sim$20\% of objects on retrograde orbits.
The circular and eccentric superearth dynamical models are very similar to each other in $q$, lacking the $q<$40~AU concentration of the control dynamical model where the $q$-distribution is dominated solely by interactions with Neptune.

These dynamical models produce clear predictions for what the orbital distributions of the high-$q$ population should look like in the absence of any observational biases. 
However, we are not able to detect all TNOs equally.  
In order to compare these models using present surveys, we must use a survey simulator (Section~\ref{sec:surveys}) to apply the known biases of the surveys to our simulated populations.

\begin{figure*}[h]
\centering
\includegraphics[scale=0.5]{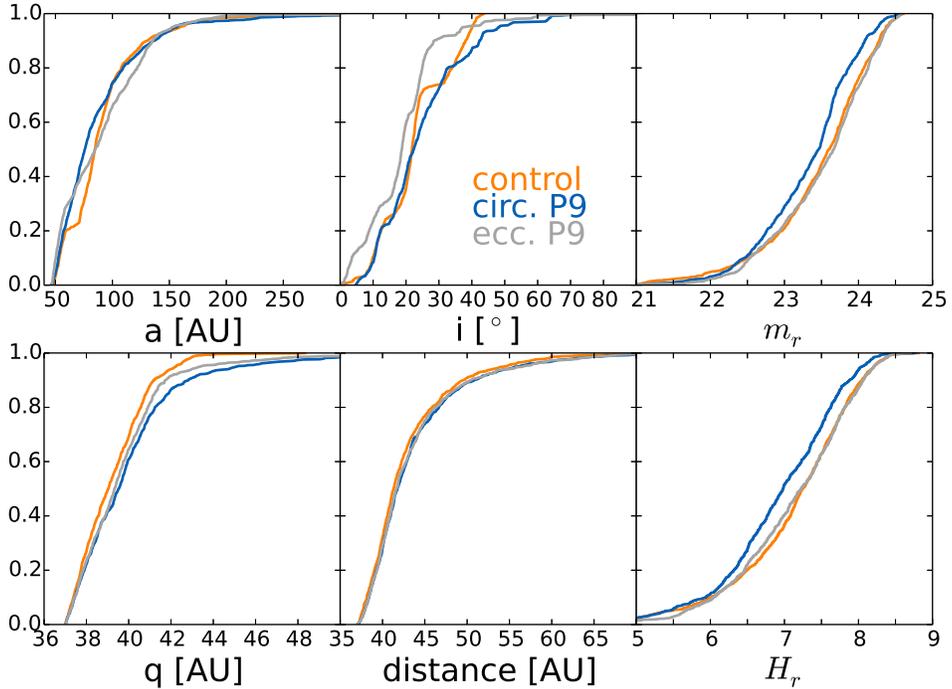}
\caption{
Cumulative histograms showing comparison between the three dynamical models biased by the survey simulator in semi-major axis $a$, inclination $i$, $r$-band magnitude $m_r$, pericenter distance $q$, distance at detection, and absolute $r$-band magnitude $H_r$.
The standard Solar System dynamical model is shown in orange, and the 9-planet dynamical models are shown in blue and gray.  
Here we use the \citet{Shankmanetal2016} divot size distribution, but a knee size distribution produces statistically and qualitatively indistinguishable results.
Despite the huge differences in intrinsic distributions (Figure~\ref{fig:cumuhist_drawnonly}), after applying survey biases, the three dynamical models are indistinguishable from each other.
}
\label{fig:cumuhist}
\end{figure*}

Figure~\ref{fig:cumuhist} shows the biased distributions in six orbital parameters.  
Immediately notable is that the three dynamical models which differ strongly are, when biased by the surveys, nearly indistinguishable from each other.  
This underscores the peril of using TNOs at the fringe of detectability and where the discovery biases are substantial and complex to assess the underlying population.
We confirmed that these biased distributions are consistent with currently published TNO detections from these surveys.

Because these surveys are all flux-limited, detectability of these objects drops sharply with distance ($d$), proportional $d^{-4}$.
Due to this effect, as $q$ increases, the probability of detection drops dramatically, and the bias towards detection of the numerous small and also lowest $q$ objects becomes overwhelming.
The flux bias effect completely overwhelms the superearth induced signature of a significant population with highly inclined orbits at high-$q$.  

Using the survey simulator, we estimate that a deep wide-field, off-ecliptic survey of several thousand square degrees, sensitive to TNOs with $i>30^{\circ}$ and $q>37$~AU, will be needed to distinguish between these dynamical models of the distant Solar System.
In our survey set, only the HiLat survey \citep[480 deg$^{2}$ to $m_g = 23.9$;][]{Petitetal2016} contained detections in this region of orbital phase space, and the three objects it provided were insufficient to make this test.
In order to debias its detections to gain absolute populations and orbital element distributions, the crucial detection and tracking efficiencies as well as all pointings of a future, deeper, high-latitude survey must be published along with the detections.  

\subsection{A Distant Ninth Planet Requires a Much Larger High-$q$ Population} \label{sec:pops}

The survey simulator draws a large number of undetectable, large $q$, TNOs before ``detecting'' the required number of simulated objects (in this case, 15, as that is the number of real detected TNOs in the four surveys inside this $a$/$q$ cut). By keeping track of the number of drawn simulated objects, we measure the absolute number of objects required by a model to produce the same number of detections as in the observed sample, down to a given $H_r$ magnitude limit. 
Using our control dynamical model with just the currently known planets, we find that the high-$q$, moderate-$a$ population for $H_r<9.0$ is $1.2\times10^5$ TNOs.  
The population required by including a circular superearth is almost three times larger, at $3\times10^5$, while an eccentric superearth requires nearly an order of magnitude larger high-$q$ population than the control simulation, at $9\times10^5$.
These population values all use the divot size distribution found to be most appropriate for the scattering population \citep{Shankmanetal2016}.
If we instead use a size distribution with a knee \citep[similar to that favored by][]{Fraseretal2014}, this approximately doubles all three required populations, while the relative population ratios remain the same.

By assuming an albedo (0.04) and density (1~g/cm$^3$), we can use the size distribution to convert the population numbers to mass estimates.  
The control dynamical model requires a present-day disk of $q>37$~AU, $50<a<500$~AU TNOs with a mass of 0.02~$M_{\oplus}$, while the circular superearth model requires a mass of 0.06~$M_{\oplus}$, and the eccentric superearth model requires a mass of 0.2~$M_{\oplus}$.
For comparison, even the control model requires a mass that is larger than the entire classical Kuiper belt \citep[0.01~$M_{\oplus}$;][]{Fraseretal2014}.  
It is important to note that current observations are rather insensitive to this high-$q$ population (Section \ref{sec:obs}), so it is unknown whether these mass estimates violate any observational constraints. 
The possibility exists that a large high-$q$ population could be hidden at the edge of observability.  

\subsection{A Distant Ninth Planet Produces no Angular Clustering} \label{sec:nocluster}

\begin{figure}[h]
\centering
\includegraphics[scale=0.5]{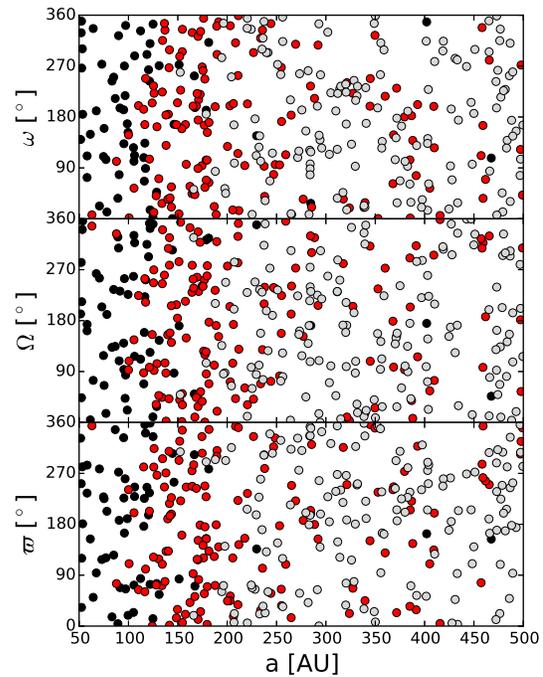}
\caption{
Distribution of argument of pericenter $\omega$, longitude of the ascending node $\Omega$, and longitude of pericenter $\varpi$ versus semimajor axis $a$ for the eccentric P9 simulation.  At the end of our simulation emplacing scattering TNOs in the presence of an eccentric ninth planet, there is no clustering of any of these angles.  This is also true for the circular P9 simulation. Points are color-coded according to pericenter distance $q$, as this dominates detectability.  The most easily detected TNOs with $q<40$~AU are black, the most difficult to detect with $q>100$~AU shown in grey, and moderate $q$ ($40<q<100$~AU) in red. This demonstrates there is also no clustering in the most easily detected low-$q$ population.
}
\label{fig:nocluster}
\end{figure}

While this work is not focused on the clustering of orbital angles originally suggested by \citet{TrujilloSheppard2014}, our dynamical simulations contain this information.  We remind the reader that this is a scattering disk that has been emplaced in the presence of a massive ninth planet.  The surviving TNOs at the end of the simulation exhibit no clustering of argument of pericenter $\omega$, longitude of pericenter $\varpi$, or longitude of the ascending node $\Omega$ (Figure~\ref{fig:nocluster}). 

While the shepherding of orbital angles has been demonstrated to be a possible dynamical effect of an eccentric, distant, massive ninth planet on a subset of TNOs by \citet{BatyginBrown2016}, their simulations do not show how strong this clustering signal is expected to be in a realistic \textit{scattering} disk.  Our simulations contain particles that do not uniformly precess, but the sample as a whole does not exhibit any clustering; this result is also seen in the work of \citet{Shankmanetal2016b}.  The N-body integrations in \citet{BatyginBrown2016} started with a flat distribution of a few hundred particles on scattering disk-like orbits, while our simulation emplaced many thousands of particles into the scattering disk and Oort Cloud in the presence of the four giant planets and a ninth planet.  Further analysis of this theory must demonstrate if (and how) the clustered TNOs are preferentially retained, as well as remove possible observational biases \citep{Shankmanetal2016b,SheppardTrujillo2016}.

\section{Discussion and Conclusion}

We find that a superearth on either a circular or eccentric orbit in the outer Solar System strongly affects the orbital distribution of the distant Kuiper belt ($q>37$~AU and $50<a<500$~AU) when compared to a control dynamical model containing only the currently known planets.
However, because flux-limited survey detections will always be dominated by the lowest $q$ objects, the strong differences between the predicted distributions are undetectable in the well-characterized surveys we examined here.

In order to match observations, the predicted mass of this high-$q$ population is 3$\times$ higher for a Solar System containing a circular superearth, and nearly 10$\times$ higher for an eccentric superearth.  
This is higher than other published estimates of the size of the population in this region, 
but we note that this high-$q$ population is not well constrained by current observations and therefore the uncertainties are large.

We do not find evidence for clustering of TNO orbital angles ($\omega$, $\Omega$, or $\varpi$) caused by either an eccentric or circular ninth planet.  Future analyses of this effect must not only demonstrate that this apparent clustering is not due merely to observational bias, but also provide an explanation for how TNOs are preferentially emplaced or retained in a portion of the ninth planet's dynamical phase space that allows this shepherding effect to dominate detected TNOs.  

The presence or absence of an additional superearth-mass planet also has important implications for the structure of the scattering TNO disk and inner Oort Cloud.  
We find that the fraction of test particles that end up in the Oort Cloud population ($q>45$~AU and $a>1000$~AU) is almost the same for each of the three surveys ($\sim$3\%),
so the presence of a distant superearth does not appear to be an important dynamical barrier to Oort Cloud production (see Figure~\ref{fig:oort}).
The fraction of objects that end up in the high-$q$ and moderate-$a$ population ($q>37$~AU, $50<a<500$~AU), however, is significantly different for the three simulations.  
With an eccentric superearth, the fraction is three times higher 
than the control simulation's value of 0.1\%
while a circular superearth produces a fraction that is nine times higher.
These population ratios are largely beyond the current realm of detectability, but could provide an important diagnostic of our Solar System's true planetary architecture in the future when compared with other TNO populations.

Using the simulations in this work, we find that a wide-field, relatively deep, off-ecliptic survey will have great power in constraining the presence or absence of an additional massive planet in our Solar System because of the widely differing inclination distributions of scattering TNOs produced by different solar system scenarios.  
This survey must be meticulous about recording detection and tracking biases, and must take care not to preferentially lose high-inclination, large $a/e$ TNOs due to tracking difficulties.
In particular, since the full $a$ and $q$ distribution contains so much
information, placing a constraint on the presence of a superearth requires tracking all large-$a$ objects to high-quality orbits; this is expensive because getting $a$ to converge for highly eccentric orbits requires many astrometric observations, over a long time period.
Making sure that the survey is sensitive to TNOs with inclinations greater than 30$^{\circ}$ and pericenters outside the immediate dynamical dominance zone of Neptune ($q\gtrsim37$~AU) is vital for distinguishing between the dynamical models presented in this work.

\acknowledgments {
The authors wish to thank an anonymous referee for suggestions which improved this manuscript.
SML gratefully acknowledges support from the NRC-Canada Plaskett Fellowship and would like to dedicate this paper to Fern May Bongarzone Lawler, born three days after manuscript submission.
This research was supported by funding from the National Research Council of Canada and the National Science and Engineering Research Council of Canada.
This research used the facilities of the Canadian Astronomy Data Centre and the Canadian Advanced Network for Astronomical Research operated by the National Research Council of Canada with the support of the Canadian Space Agency.
}

\bibliographystyle{astron}

\end{document}